# Dislocation interactions during low-temperature plasticity of olivine strengthen the lithospheric mantle

David Wallis[1], Lars. N. Hansen[2,3], Kathryn M. Kumamoto[2], Christopher A. Thom[2], Oliver Plümper[1], Markus Ohl[1], William B. Durham[4], David L. Goldsby[5], David E.J. Armstrong[6], Cameron D. Meyers[3,7], Rellie Goddard[2], Jessica M. Warren[8], Thomas Breithaupt[2], Martyn R. Drury[1], Angus J. Wilkinson[6]

[1]Department of Earth Sciences, Utrecht University, Utrecht, 3584 CB, The Netherlands.
[2]Department of Earth Sciences, University of Oxford, Oxford, OX1 3AN, U.K.
[3]Department of Earth Science, University of Minnesota-Twin Cities, Minneapolis, Minnesota, U.S.A. 55455.
[4]Department of Earth, Atmospheric and Planetary Sciences, Massachusetts Institute of Technology, Cambridge, Massachusetts, 02139-4307, U.S.A.
[5]Department of Earth and Environmental Science, University of Pennsylvania, Philadelphia, Pennsylvania, 19104, U.S.A.
[6]Department of Materials, University of Oxford, Oxford, OX1 3PH, U.K.
[7]Department of Earth, Environmental, and Planetary Sciences, Brown University, Providence, Rhode Island, 02912, U.S.A.
[8]Department of Geological Sciences, University of Delaware, Newark, Delaware, 19716, U.S.A.

## Abstract

The strength of the lithosphere is typically modelled based on constitutive equations for steady-state flow. However, models of lithospheric flexure reveal differences in lithospheric strength that are difficult to reconcile based on such flow laws. Recent rheological data from low-temperature deformation experiments on olivine suggest that this discrepancy may be largely explained by strain hardening. Details of the mechanical data, specifically the effects of temperature-independent back stresses stored in the samples, indicate that strain hardening in olivine occurs primarily due to long-range elastic interactions between dislocations. These interpretations provided the basis for a new flow law that incorporates hardening by development of back stress. Here, we test this hypothesis by examining the microstructures of olivine samples deformed plastically at room temperature either in a deformation-DIA apparatus at differential stresses of ≤ 4.3 GPa or in a nanoindenter at applied contact stresses of ≥ 10.2 GPa. High-angular resolution electron backscatter diffraction maps reveal the presence of geometrically necessary dislocations with densities commonly above $10^{14}$ m$^{-2}$ and intragranular heterogeneities in residual stress on the order of 1 GPa in both sets of samples. Scanning transmission electron micrographs reveal straight dislocations aligned along slip bands and interacting with dislocations of other types that act as obstacles. The stress heterogeneities and accumulations of dislocations along their slip planes are consistent with strain hardening resulting from long-range back-stresses acting between dislocations. These results corroborate the mechanical data in supporting the form of the new flow law for low-temperature plasticity and provide new microstructural criteria for identifying the operation of this deformation mechanism in natural samples. Furthermore, similarities in the structure and stress fields of slip bands formed in single crystals deformed at low temperatures and those formed at high temperatures suggest that similar hardening processes occur in both regimes, providing a new constraint for models of transient creep at high temperatures.

# 1 Introduction

Determining the strength of lithospheric plates is a key objective in geodynamics and tectonics. The strength of the lithosphere is incorporated into models of processes including global convection (van Heck and Tackley, 2008), deformation in collision zones (England and Houseman, 1986; England and Molnar, 2015), and plate flexure beneath surface loads (Zhong and Watts, 2013) and in subduction zones (Buffett and Becker, 2012; Hunter and Watts, 2016). Accordingly, the strength of the lithosphere is determined based on observations from each tectonic setting (e.g., Watts et al., 2013) and characterised in the form of rheological models with general applicability based on experimentally derived flow laws (e.g., Burov, 2011; Kohlstedt et al., 1995; Watts et al., 2013). These approaches indicate that in many settings the maximum stress that can be supported in a lithospheric section, and therefore much of the integrated strength, is likely controlled by low-temperature plasticity (Buffett and Becker, 2012; England and Molnar, 2015; Hansen et al., 2019; Hunter and Watts, 2016; Mei et al., 2010; Zhong and Watts, 2013).

Geophysical observations indicate that the strength of the lithosphere varies between tectonic settings. One proxy for lithospheric strength that provides a convenient metric to compare between regions is the elastic thickness (Walcott, 1970). Compilations of elastic thicknesses indicate that strength varies by an order of magnitude due to lithospheric structure and thermal regime in a manner broadly consistent with predictions from flow laws for steady-state deformation (Burov, 2011; Watts et al., 2013). However, recent models of flexure data have revealed differences in lithospheric strength that are difficult to reconcile based on conventional flow laws. For instance, lithospheric-strength profiles incorporating the flow law of Mei et al. (2010) for low-temperature plasticity of olivine provide good fits in models of continental shortening in the Tien Shan (England and Molnar, 2015) and flexure of the Pacific plate at subduction zones (Hunter and Watts, 2016). However, Zhong and Watts (2013) found that similar profiles, also based on the flow law of Mei et al. (2010), overpredicted the strength of the lithosphere in models of flexure of the Pacific plate around Hawaiian seamounts. They found that the pre-exponential term in the flow law of Mei et al. (2010) must be reduced by a factor of $10^6$–$10^8$ to reproduce the observed flexural response. Hunter and Watts (2016) suggested that subducting lithosphere is apparently strengthened as a result of unrelaxed viscoelastic stresses or that lithosphere at Hawaii is apparently weakened as a result of either bending stresses from plate cooling (Wessel, 1992) or magma assisted flexure (Buck et al., 2015).

Recent laboratory experiments have led to a new hypothesis regarding the cause of differences in lithospheric strength. Hansen et al. (2019) investigated low-temperature plasticity of single crystals and aggregates of olivine through experiments in a deformation-DIA (D-DIA) apparatus. Experiments were conducted mostly at room temperature to inhibit processes involving long-range diffusion, such as diffusion creep and dislocation climb. Both single crystals and aggregates exhibited strain hardening of 1–2 GPa between the yield stress and flow stress. Importantly, Hansen et al. (2019) conducted experiments in which the samples were subjected to cycles of shortening and extension. These experiments revealed that the yield stress during the initial loading was consistently greater in magnitude than those on subsequent cycles. This behaviour, termed the Bauschinger effect, is widely recognised in the metallurgical literature and indicates that strain hardening arises, at least in part, from long-range elastic interactions among dislocations, termed kinematic hardening (e.g., Kuhlmann-Wilsdorf and Laird, 1979). These interactions manifest as back stresses among dislocations that reduce the apparent yield stress upon load reversal. Based on these mechanical data, Hansen et al. (2019) formulated a new flow law for low-temperature plasticity in olivine that characterises strain hardening by accumulation of back stress. Lithospheric-strength profiles for 80

Myr-old oceanic lithosphere, that incorporate the new flow law and are constructed for a range of strains, indicate that integrated lithospheric strength can potentially double over the first 2% strain, varying between values broadly consistent with those required in the models of flexure at Hawaiian seamounts at low strains and Pacific subduction zones at higher strains (Hansen et al., 2019). These results indicate that hardening by the accumulation of back stress with plastic strain may be an important consideration for models of lithospheric strength. This hypothesis is related to the concept of unrelaxed viscoelastic stresses suggested by Hunter and Watts (2016) but additionally specifies the microphysical processes involved, at least in the portions of the lithosphere deforming by low-temperature plasticity.

Whilst clues to the microphysical mechanisms of strain hardening in olivine can be gleaned from mechanical data, the key processes are yet to be characterised in detail based on microstructural observations. The portions of stress-strain curves that express strain hardening are broadly similar in shape across temperatures spanning 25–1500°C but hardening typically occurs over longer strain intervals and is greater in magnitude at lower temperatures (Chopra, 1997; Cooper et al., 2016; Druiventak et al., 2011; Hansen et al., 2019; Hanson and Spetzler, 1994; Post, 1977). The similarity between strain hardening in single crystals and aggregates across this temperature range indicates that intragranular processes contribute to hardening in all cases (Chopra, 1997; Hansen et al., 2019; Hanson and Spetzler, 1994). Hanson and Spetzler (1994) demonstrated that hardening of single crystals at a temperature of 1377°C is controlled by increases in dislocation density. In aggregates, strain incompatibility among mechanically anisotropic grains potentially makes an additional contribution to hardening (Ashby and Duval, 1985). Yet, the causes of hardening at lower temperatures are more poorly constrained. Thieme et al. (2018) did not resolve changes in intragranular misorientation during strain hardening of aggregates at temperatures of 1000–1200°C using conventional electron backscatter diffraction (EBSD). However, transmission electron microscope (TEM) images of olivine deformed at 600–1000°C reveal cellular structures of tangled dislocations, inferred to contribute to strain hardening (Phakey et al., 1972; Druiventak et al., 2011). Recently, Kumamoto et al. (2017) used high-angular resolution electron backscatter diffraction (HR-EBSD) to detect densities of geometrically necessary dislocations (GNDs) on the order of $10^{14}$ m$^{-2}$ around indents made at room temperature in single crystals of olivine. They inferred that interactions among the dislocations played a key role in strengthening the material. However, as the elastic analysis of Durinck et al. (2007) suggests that short-range interactions (e.g., formation of junctions) between dislocations may be weak in olivine, and previous studies have not mapped the stress fields associated with long-range elastic interactions, the contributions of these strengthening mechanisms to strain hardening remain unclear.

In the present paper, we analyse the microstructures of samples of olivine deformed by low-temperature plasticity to test the hypothesis that strain hardening results primarily from long-range elastic interactions between dislocations. We utilise samples deformed at room temperature in a D-DIA apparatus by Hansen et al. (2019) and in nanoindentation experiments by Kumamoto et al. (2017) to assess the processes operating in both sets of experiments. We employ HR-EBSD to simultaneously map densities of GNDs and heterogeneity in residual stress, along with scanning transmission electron microscope (STEM) images of dislocation arrangements. The results provide critical new constraints for models of strain hardening in the lithospheric mantle.

# 2 Methods

## *2.1 Deformation-DIA experiments*

We analyse the microstructures of two starting materials and four deformed samples from the experiments of Hansen et al. (2019). These experiments were conducted in a D-DIA apparatus housed on the 6-BM-B beamline of the Advanced Photon Source at Argonne National Laboratory. A synchrotron X-ray source was employed to measure bulk strain through radiography and to determine elastic strain, and thus the stress state, through energy-dispersive X-ray diffraction. The experiments were conducted at temperatures of 25–1200°C, pressures of 0.2–10.4 GPa, and differential stresses up to 4.3 GPa. Here, we focus on samples deformed at room temperature as this condition effectively excludes temperature-dependent processes. Microstructures formed at room temperature have not been characterised in detail by previous studies.

Both starting materials that we analyse, PT-1184 and PI-1519, were fabricated by hot-pressing powdered San Carlos olivine at 1250°C in a gas-medium apparatus. Sample PI-1519 was also deformed at 1100–1250°C as part of the data set presented by Hansen et al. (2011), prior to use in the D-DIA experiments. The grain sizes of samples PT-1184 and PI-1519, measured by the line-intercept method using EBSD maps, are 42.9 µm and 4.6 µm, respectively (Hansen et al., 2019), including a scaling factor of 1.5 (Underwood, 1970 pp. 80–93).

Importantly, Hansen et al. (2019) performed an annealing stage of approximately 10 minutes, at a temperature of 1000°C and a pressure of several GPa, at the start of each experiment, which served to relax the macroscopic differential stress and internal stress heterogeneity imparted during initial pressurisation. These stresses potentially arise from the slight non-cubic geometry of the assembly, elastic anisotropy in crystalline aggregates, and the stress fields of dislocations. Figure 1 presents examples of the effect of this annealing step on peak width in the X-ray diffraction patterns, which is a proxy for internal stress heterogeneity. These data were collected during recent experiments that followed the same procedure as those of Hansen et al. (2019). Figure 1a demonstrates the contrast between a diffraction pattern from a highly stressed sample and one collected at the end of the annealing treatment. The annealed sample exhibits narrower peaks and a more distinct doublet of the {112} and {131} peaks than the stressed sample. Figure 1b demonstrates that differential stresses generated during initial pressurisation are fully relaxed before the start of the experiments. Figure 1c documents the reduction in the widths of diffraction peaks to approximately constant values during the annealing stage, indicating relaxation of pre-existing stress heterogeneity within the samples. We collected HR-EBSD data from portions of PT-1184 and PI-1519 extracted prior to this annealing step and therefore the stress heterogeneities and GND densities present in those datasets provide upper bounds on those actually present at the beginning of each D-DIA experiment.

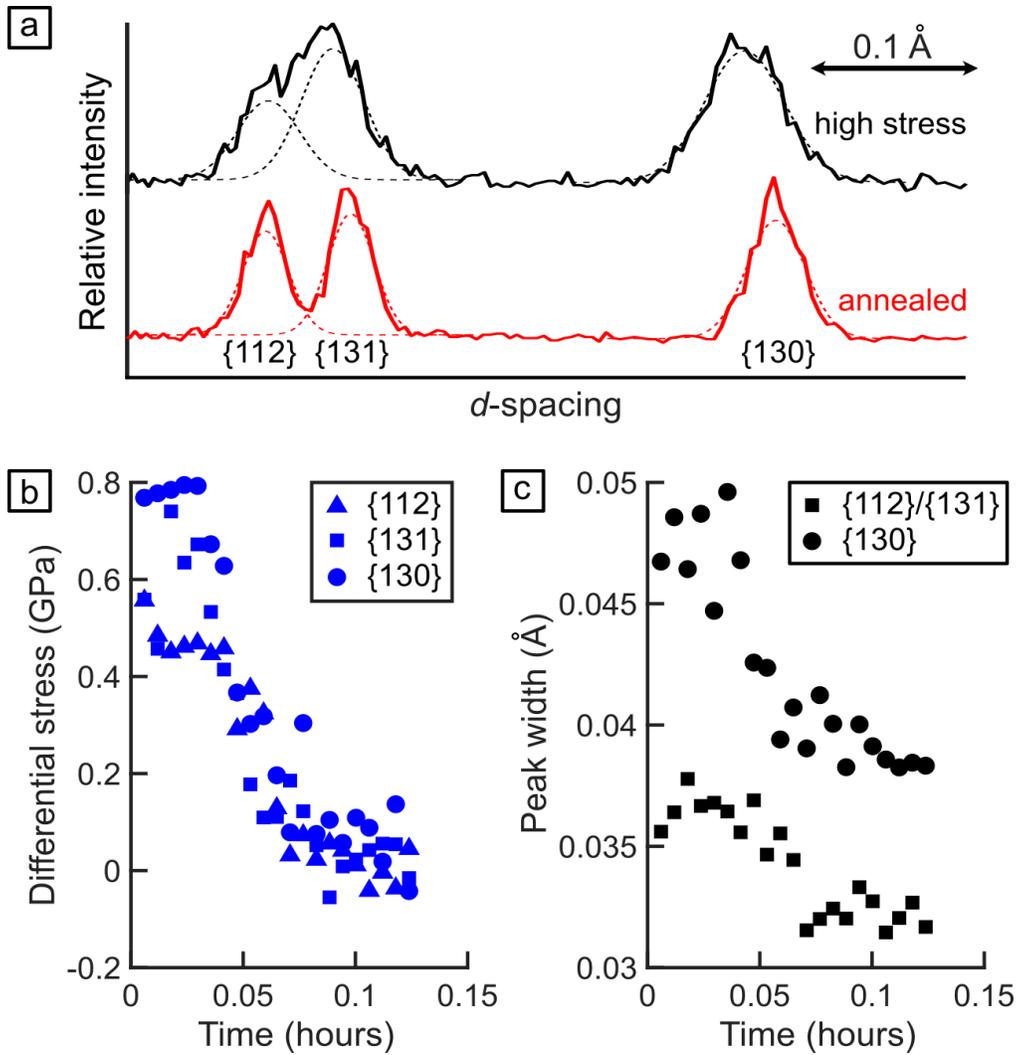

*Figure 1.* Examples of the effects of annealing on X-ray diffraction patterns and macroscopic differential stresses. (a) Example diffraction patterns demonstrating the differences in peak width from a highly stressed sample and a sample in which internal stress has been reduced by annealing at 1000°C at the beginning of a D-DIA experiment. (b) Differential stresses calculated from X-ray diffraction measurements of the d-spacings of three lattice planes, {hkl}, during annealing at 1000°C at the beginning of a D-DIA experiment (San432). (c) Full width at half maximum of the peaks corresponding to the same three lattice planes. The widths of the {112} and the {131} peaks have been set as equal in the initial peak fitting as they form the doublet evident in (a) and are not always separable.

The experiments that we analyse, San377 and San382, both consisted of assemblies of two samples, with one stacked on top of the other and deformed simultaneously. We add the suffixes 'b' and 't' to the experiment names given by Hansen et al. (2019) to indicate samples that were located on the bottom and top of their assemblies, respectively. The experimental conditions and key data are summarised in Table 1. Sample San377b was shortened to a strain of 18%. Sample San382 was deformed in an experiment that imposed shortening, followed by extension, and then final shortening. During these cycles, the maximum strains imposed on San382t and San382b were 13.5% and 8.4%, respectively.

*Table 1* Summary of deformation experiments and EBSD maps

| Sample | Experiment | Grain size (μm) | EBSD map size (pixels) | EBSD step size (μm) | Notes |
|---|---|---|---|---|---|
| MN1 | Undeformed | > 1000 | 500 × 400 | 0.5 | Single crystal typical of starting material for D-DIA and nanoindentation experiments |
| PT-1184 | Undeformed | 42.9 | 400 × 316 | 0.5 | Isostatically hot pressed starting material for D-DIA experiments |
| PI-1519 | Paterson | 4.6 | 800 × 600 | 0.2 | Starting material for D-DIA experiments, previously deformed at temperatures of 1373–1523 K |
| San382 | D-DIA | ~700 and 3.0 | 2097 × 2051 | 0.5 | Paired sample deformed at a temperature of 298 K, pressures of 5.3–8.2 GPa, and differential stresses of -3.8–4.3 GPa. |
| San382t | D-DIA | ~700 | 295 × 155 | 0.15 | Single crystal portion of San382 |
| San382b | D-DIA | 3.0 | 320 × 200 | 0.15 | Aggregate portion of San382 |
| San377b | D-DIA | 42.9 | - | - | Part of a paired sample deformed at a temperature of 298 K, pressures of 5.6–10.1 GPa, and differential stresses of < 4.3 GPa. |
| OP4-2 300 nm | Nanoindentation | > 1000 | 90 × 90 | 0.1 | Indent to a maximum depth of 300 nm in an undeformed single crystal. Yield stress was 10.5 GPa. |
| OP4-2 800 nm | Nanoindentation | > 1000 | 210 × 210 | 0.1 | Indent to a maximum depth of 800 nm in an undeformed single crystal. Yield stress was 10.8 GPa. |
| PI-1488 600 nm | Nanoindentation | 9.6 | - | - | Indent to a maximum depth of 600 nm in an aggregate previously deformed at temperatures of 1373–1423 K. Yield stress was 10.2 GPa. |

Sections of the samples deformed in D-DIA experiments were prepared parallel to the loading column and near the centre of each sample. Both these samples and the starting materials were polished with successively finer diamond grits down to a grit size of 0.05 μm. The polished surfaces were coated with 0.5 nm of Pt/Pd for EBSD mapping and 8 nm of Pt/Pd for focused ion beam scanning electron microscopy (FIB-SEM).

## *2.2 Nanoindentation experiments*

We analyse the microstructures of three indents from the experiments of Kumamoto et al. (2017). During nanoindentation experiments, dislocation motion can be activated at low temperatures due to confinement imposed by the sample material and indenter, which inhibits fracturing during loading (Swain and Hagan, 1976). Kumamoto et al. (2017) performed nanoindentation experiments at room temperature on an MTS Nanoindenter XP in the Department of Materials, University of Oxford, using a conospherical diamond tip with an effective radius of 3.0 μm. Samples were polished with successively finer diamond grits and finished with 0.03 μm colloidal silica prior to nanoindentation. The maximum depths reached for the indents in OP4-2 were 300 nm and 800 nm, whereas the maximum depth reached for the indent in PI-1488 was 600 nm.

The starting materials for the nanoindentation experiments consisted of San Carlos olivine. Two of the indents that we analyse were placed in a gem-quality single crystal, sample OP4-2. The other indent was in an aggregate, sample PI-1488, that was statically hot-pressed and then deformed at a final temperature of 1150°C by Hansen et al. (2011).

## *2.3 Electron backscatter diffraction*

EBSD data were acquired on field-emission gun SEMs at the University of Oxford, Utrecht University, and the Zeiss Microscopy Centre. All three instruments were equipped with Oxford Instruments Nordlys Nano or Symmetry detectors and AZtec acquisition software. Map dimensions are presented in Table 1. Reference-frame conventions were validated following the approach of Britton et al. (2016). For all HR-EBSD maps presented here, $x_1$ is horizontal, $x_2$ is vertical, and $x_3$ is out of the plane of the map. For HR-EBSD analysis, diffraction patterns were saved at a resolution of $1344 \times 1024$ pixels. Cross-correlation-based postprocessing was performed following the method of Wilkinson et al. (2006) and Britton and Wilkinson (2012, 2011). The cross-correlation procedure maps the shifts in 100 regions of interest within the diffraction patterns relative to their positions in a reference pattern within each grain to overdetermine the deformation gradient tensor describing lattice rotations and elastic strains. Densities of geometrically necessary dislocations were calculated from the rotation fields following the method of Wallis et al. (2016). As the appropriate dislocation types to fit to the lattice curvature generated at low temperatures are not clear *a priori*, we present maps of the total dislocation density, which should be affected little by potential inaccuracies in the dislocation types used to fit the lattice curvature. Stress heterogeneities were calculated from the strain heterogeneities using Hooke's law and the elastic constants for olivine at room temperature and a confining pressure of 1 atm (Abramson et al., 1997).

Measured elastic strains and residual stresses are relative to the strain and stress state at the reference point in each grain. Reference points were chosen in areas of good pattern quality away from grain boundaries. In the maps of the indents, regions far from the indents are likely free from significant elastic strain, and therefore HR-EBSD provides maps of the absolute residual stresses around the indents. In contrast, no portion of the samples deformed in D-DIA experiments can be assumed to be strain-free. Therefore, the stress data from these samples were normalised by subtracting the mean of each component of the measured stress tensor within each grain, providing maps of relative stress heterogeneity (Wallis et al., 2019, 2017).

## 2.4 Focused ion beam and transmission electron microscopy

Electron-transparent foils were prepared using an FEI Helios Nanolab G3 Dualbeam FIB-SEM at Utrecht University. For samples deformed in D-DIA experiments, foils were oriented such that the plane of each foil was normal to the maximum compression direction. For samples deformed in nanoindentation experiments, foils were oriented parallel to the indentation direction. Transmission electron microscope (TEM) investigations of the foils were carried out with an FEI Talos F200X operated at 200 kV. Bright-field and dark-field images were acquired simultaneously in STEM mode.

# 3 Results

## 3.1 Starting materials

Figure 2 presents HR-EBSD maps of samples PT-1184 and PI-1519, typical of those used as starting materials for D-DIA experiments. Maps of lattice rotations reveal subgrains as misoriented intragranular domains. The boundaries of these domains are evident as bands of elevated GND density representing subgrain boundaries. In the isostatically hot-pressed sample, PT-1184, apparent dislocation densities in subgrain interiors are on the order of $10^{12}$ m$^{-2}$, which corresponds to the noise level (Wallis et al., 2016). In contrast, subgrain interiors in sample PI-1519, deformed at temperatures of 1373–1523 K by Hansen et al. (2011), contain elevated GND densities on the order of $10^{14}$ m$^{-2}$. Similarly, intragranular stress heterogeneities in PT-1184 have magnitudes on the order of a few hundred megapascals, whereas those in PI-1519 have magnitudes on the order of 1 GPa.

## 3.2 Samples deformed in deformation-DIA experiments

### 3.2.1 Electron backscatter diffraction

Figure 3 presents EBSD data from experiment San382, which consists of a single crystal (San382t) stacked on top of an aggregate with a grain size of 3.0 μm (San382b). The single crystal was loaded in approximately the [111]$_c$ orientation, following the convention of Durham and Goetze (1977). Figure 3a presents a map coloured by the misorientation axis between the orientation at each pixel and the average orientation of each grain. This colour scheme reveals conjugate sets of diagonal bands with spacings on the order of tens of micrometers. A map based on the contrast of bands in the electron diffraction patterns (Figure 3d) reveals bands with finer spacings of a few micrometers. A profile of misorientation angles across the single crystal (Figure 3e) demonstrates that these banded structures are associated with misorientation angles up to a few degrees. Orientations within the single crystal are dispersed around [010] (Figure 3b). Pole figures of crystal orientations within the aggregate (Figure 3c) exhibit a maximum of [010] axes aligned with the compression direction, whereas [100] and [001] axes both exhibit two weaker maxima perpendicular to the compression direction. This distribution of crystal orientations contrasts with that of the starting material (Figure 3c), which lacked a preferred orientation.

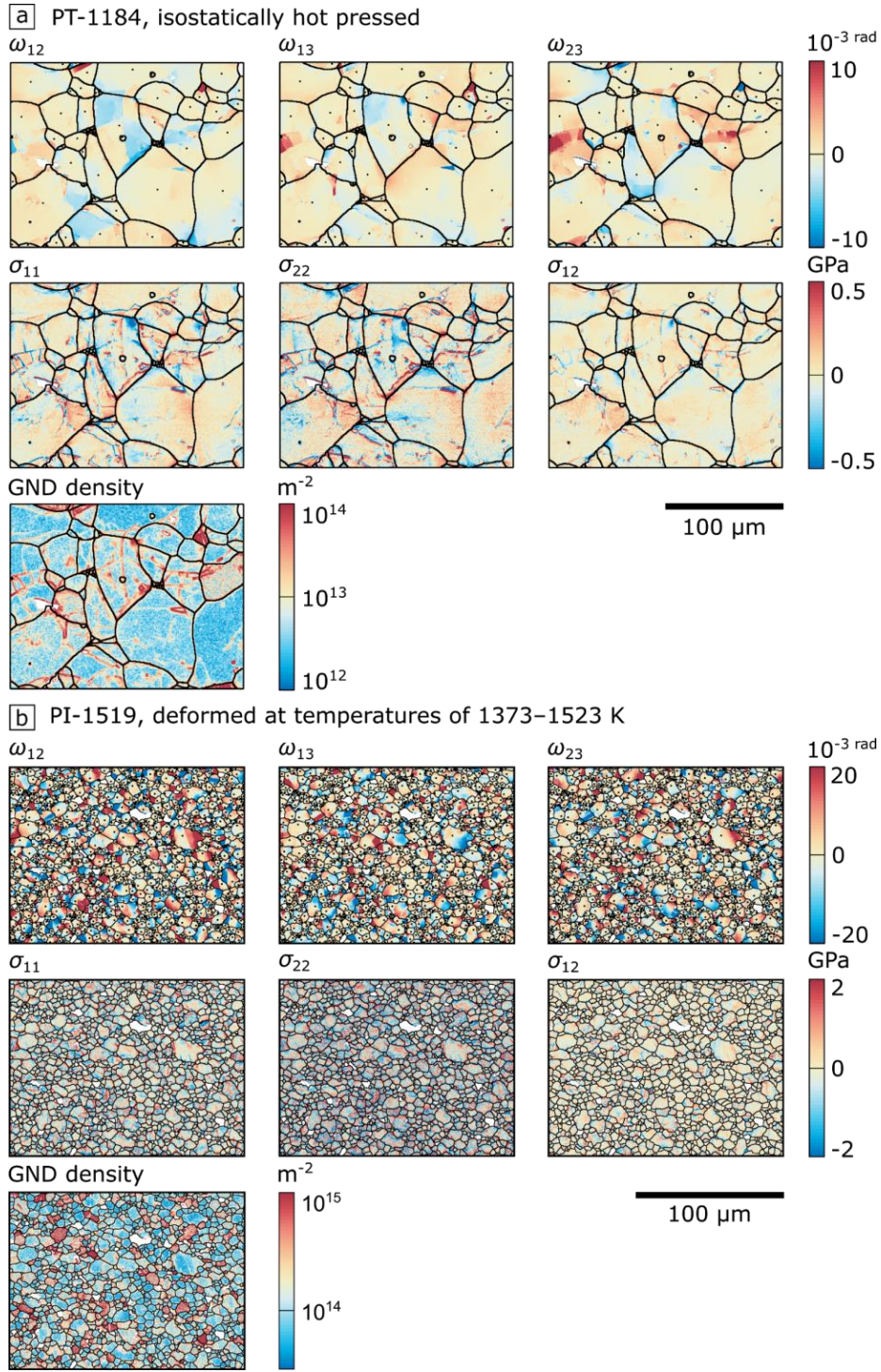

*Figure 2.* HR-EBSD maps of lattice rotations ($\omega_{ij}$), residual stress heterogeneity ($\sigma_{ij}$), and GND density in starting materials for D-DIA experiments. (a) Sample PT-1184, which was isostatically hot pressed but not deformed. (b) Sample PI-1519, which was isostatically hot pressed and then deformed at temperatures of 1373–1523 K by Hansen et al. (2011). Lattice rotations are relative to the crystal orientations at the reference points marked with black dots in each grain. Components of the residual stress tensor ($\sigma_{ij}$) are normalised to the mean of each component within each grain.

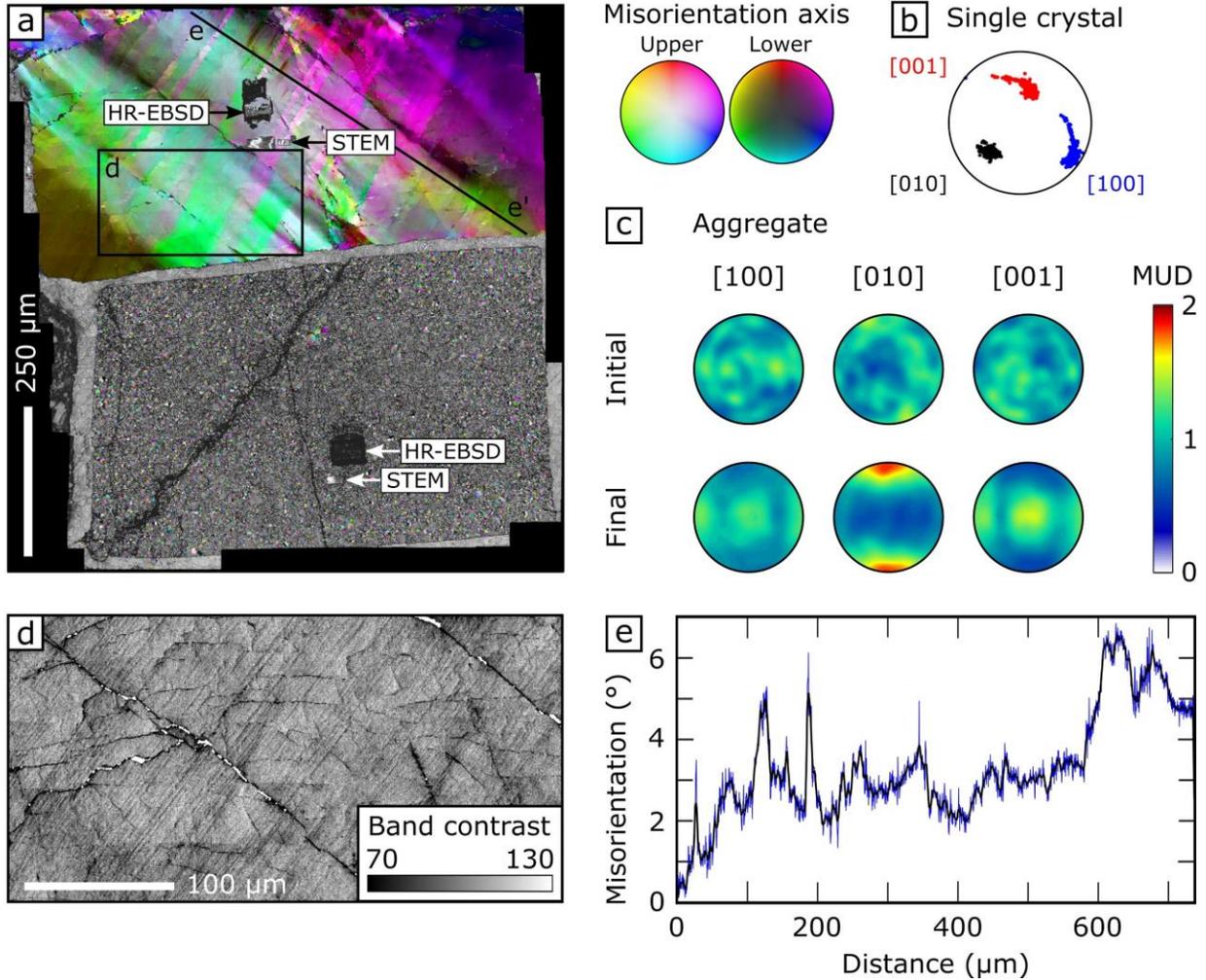

*Figure 3.* EBSD data from experiment San382, which consists of a single crystal (San382t) stacked on top of an aggregate with a grain size of 3.0 µm (San382b). The initial compression direction is vertical. a) Overview EBSD map with olivine coloured by the misorientation axis between each point and the mean orientation of the grain. Areas not indexed as olivine are presented as greyscale band contrast. Annotations mark the areas where HR-EBSD data and liftouts for STEM imaging were acquired. b) Pole figure presenting 50,000 randomly selected orientation measurements from the single crystal in the reference frame of the map in (a). c) Pole figures of CPO in the starting material (sample 33, marked "Initial") and deformed aggregate (marked "Final") in the reference frame of the map in (a). Colours indicate multiples of uniform distribution (MUD). d) Band contrast map of the area marked in a). e) Profile of misorientation angles relative to the first point along the profile marked e–e' in a).

### 3.2.2 High-angular resolution electron backscatter diffraction

Figure 4 presents HR-EBSD maps of samples San382t and San382b. The single crystal (San382t, Figure 4a) exhibits linear domains, on the order of 1 µm in apparent width, in which the lattice is rotated on the order of $5\times10^{-3}$ radians (~0.3°) between domains. The sign of the lattice rotation alternates between domains. We refer to this microstructure as corrugated crystal lattice. The boundaries of each domain are marked by bands of elevated GND density, on the order of $10^{14}$ m$^{-2}$. Maps of residual stress heterogeneity exhibit structure broadly similar to that of the lattice rotations. The $\sigma_{11}$, $\sigma_{22}$, and $\sigma_{12}$ components vary in magnitude by ~1 GPa, and alternate in sign, over distances of approximately 1 µm.

The aggregate (San382b, Figure 4b) exhibits even greater intragranular lattice distortion than the single crystal. Lattice rotations are typically on the order of $10^{-2}$ radians and GND densities commonly approach $10^{15}$ m$^{-2}$. Intragranular heterogeneities in the in-plane normal stresses typically have magnitudes of several gigapascals. Intragranular distortion within the aggregate lacks the linear, periodic structure evident in the single crystal.

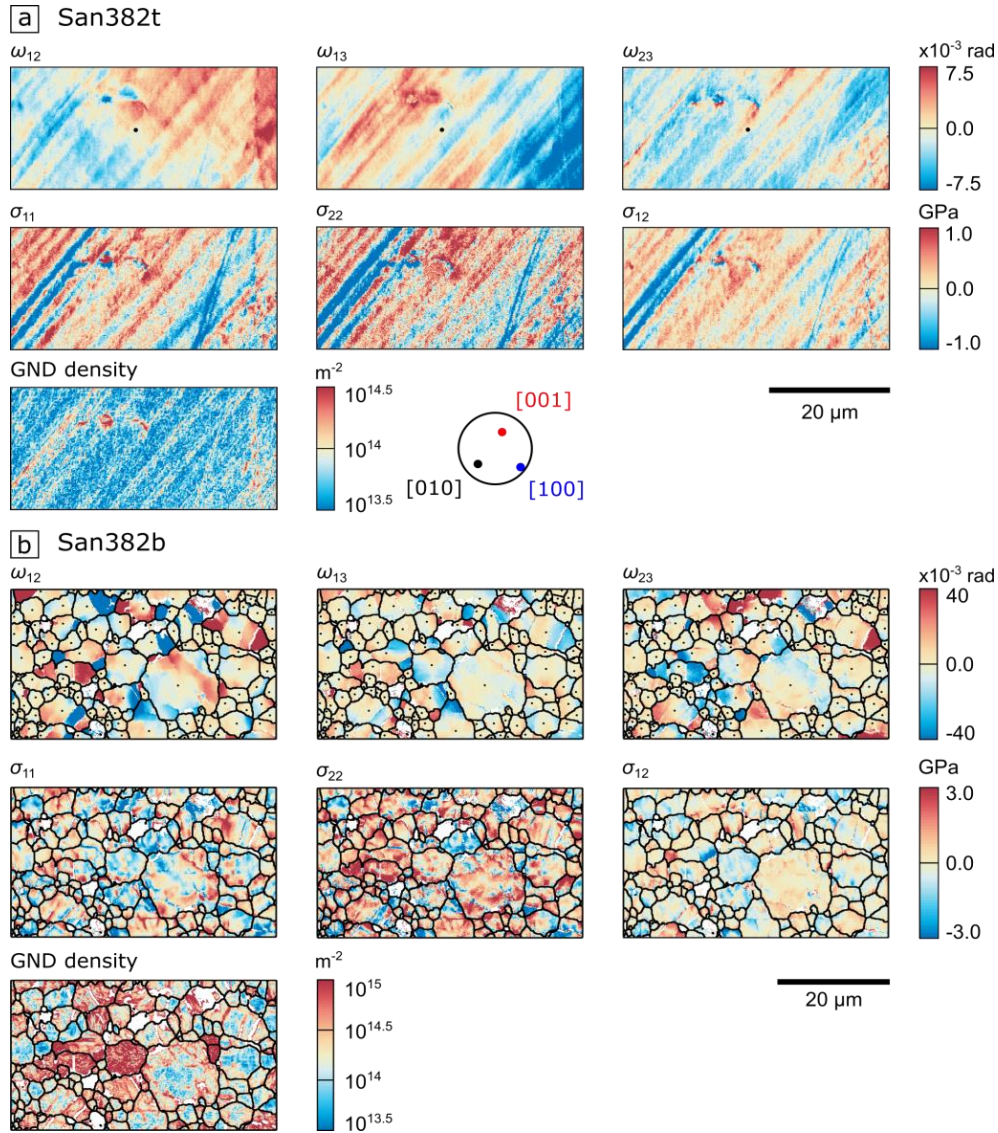

*Figure 4.* HR-EBSD maps of lattice rotations ($\omega_{ij}$), residual stress heterogeneity ($\sigma_{ij}$), and GND density in the single-crystal sample San382t (a) and aggregate San382b (b). The orientation of the single crystal in (a) is the same as presented in Figure 3b. Rotations are relative to the crystal orientation at the reference points marked with black dots. Components of the residual stress tensor are normalised to the mean of each component within each grain. The locations of the HR-EBSD maps are marked on the overview EBSD map in Figure 3a.

Figure 5 presents distributions of in-plane shear stress, $\sigma_{12}$, in undeformed and deformed single crystals and aggregates. The crystal-plasticity finite-element model predictions of Kartal et al. (2015) suggest that $\sigma_{12}$ is the component of the stress tensor least modified by stress relaxation during sectioning. The undeformed single crystal (MN1) and undeformed aggregate (PT-1184, Figure 2a) exhibit similar distributions, with standard deviations of 70 MPa and 140 MPa, respectively. The deformed single crystal (San382t, Figure 4a) contains greater stress heterogeneity than the undeformed samples, with a standard deviation of 400 MPa. The deformed aggregate (San382b, Figure 4b) displays the greatest intragranular stress heterogeneity, with a standard deviation of 1.14 GPa. As this sample has the smallest grain size and the greatest intragranular stress heterogeneity, it also has the greatest spatial gradients in stress.

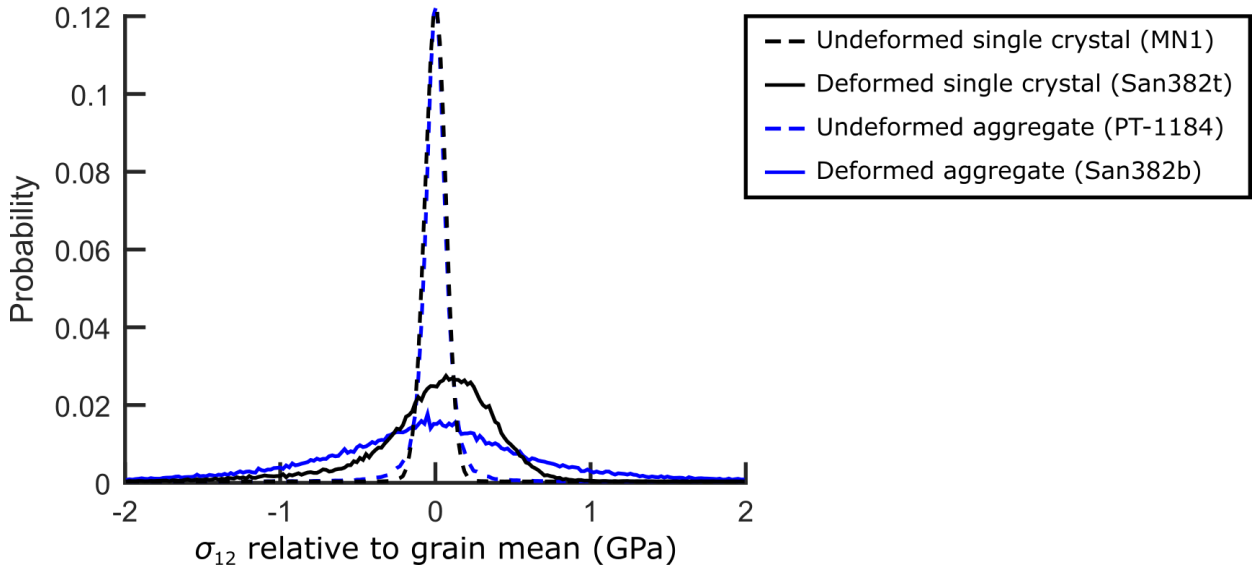

*Figure 5.* Intragranular heterogeneity in the $\sigma_{12}$ component of the residual stress tensor in undeformed and deformed single crystals and aggregates of olivine. Stresses are presented relative to the mean of the $\sigma_{12}$ component within each grain.

*3.2.3 Scanning transmission electron microscopy*

Figures 6a–c present STEM images of single crystal San382t. Dislocations in this sample form two perpendicular sets of structures trending along the diagonals of the micrographs, resulting in a cellular substructure (Figures 6a and 6b). Both sets of structures consist of planar arrays of dislocations (Figures 6b and 6c). Structures in which individual dislocations are visible reveal dislocations regularly spaced at intervals of < 100 nm (Figure 6c). The characteristics (e.g., shape, diffraction contrast) of the structures are commonly different across the points at which the structures intersect, indicating changes in the densities or types of their constituent dislocations.

Figures 6d–f present STEM images of one grain in the coarse-grained aggregate San377t. Dislocations are homogeneously distributed and spaced approximately 100 nm apart. Dislocations typically contain straight segments a few hundred micrometers in length (Figures 6d and 6f). Straight segments are often separated by shorter linking stepovers or curved segments (Figures 6d and 6f), which, at their most dense, generate more chaotic dislocation networks and heterogeneous densities (Figure 6e).

Figures 6g–i present STEM images of the fine-grained aggregate San382b. As in the other two samples, dislocations are typically straight or gently curved, with occasional curved linking segments forming open networks. Elevated dislocation densities (Figure 6g) and dislocation sources (below the bright pore in Figure 6i) are present near grain boundaries. Diffraction-contrast fringes suggest that dislocations are split into partial dislocations bounding stacking faults (Figure 6h).

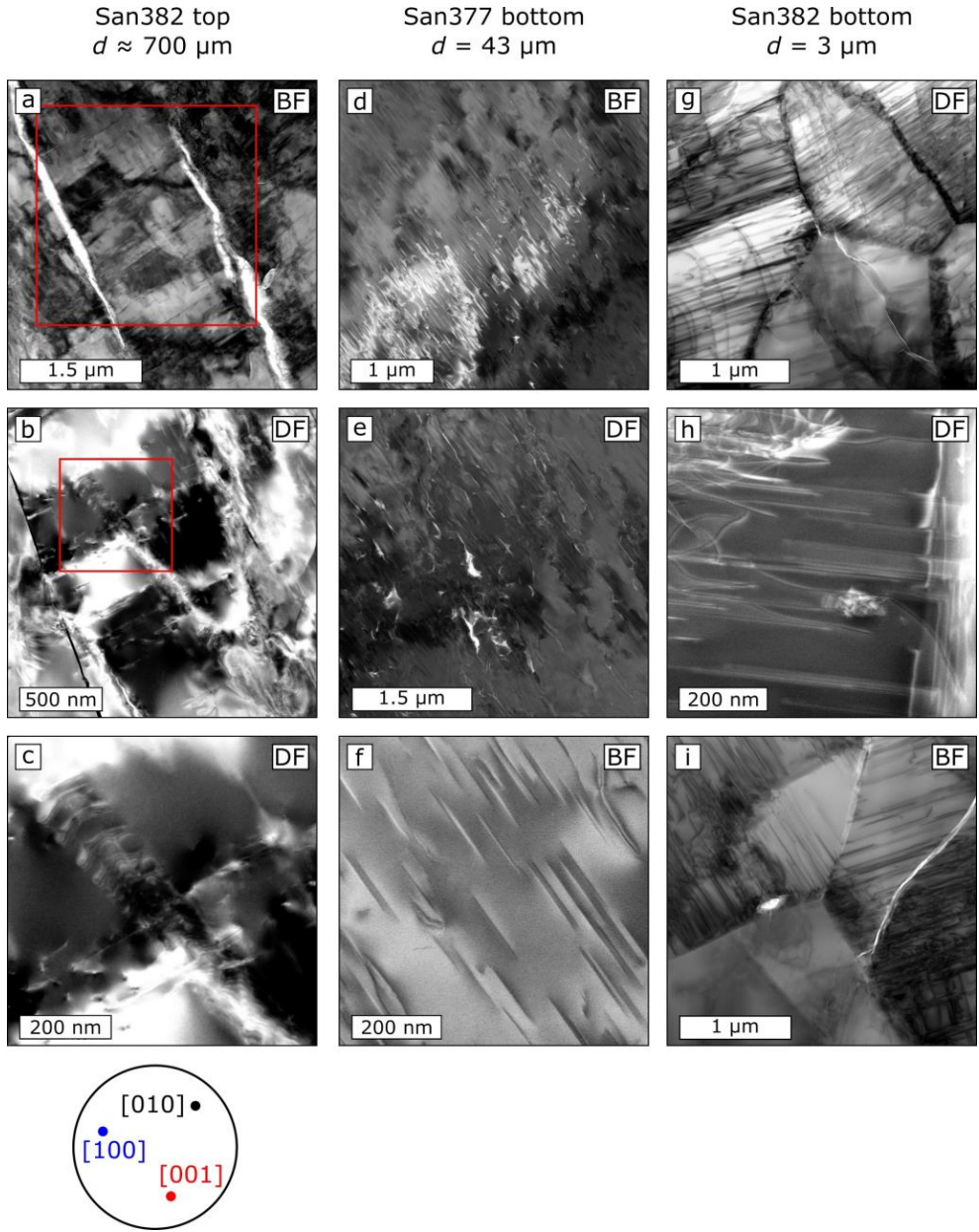

*Figure 6. STEM images of samples deformed in D-DIA experiments. Sample names and grain sizes are given for each row. (a–c) and (g–i) correspond to the same samples for which HR-EBSD maps are presented in Figure 4a and 4b, respectively, and the locations of the liftouts are marked on the overview EBSD map in Figure 3a. Red boxes in (a) and (b) indicate the areas of (b) and (c), respectively. The pole figure indicates the crystal orientation in (a–c). BF and DF indicate images taken in bright-field and dark-field modes, respectively.*

## *3.3 Samples deformed in nanoindentation experiments*

### *3.3.1 High-angular resolution electron backscatter diffraction*

Figure 7 presents HR-EBSD maps of nanoindents in the single crystal of olivine, OP4-2. The 300-nm indent in Figure 7a and the 800-nm indent in Figure 7b exhibit broad similarities in the distributions of lattice rotation, GND density, and residual stress around the indents. However, the magnitudes of these parameters and the size of the deformed zone are greater around the 800 nm indent. Sharp discontinuities in the rotation fields around the 800-nm indent mark the traces of microcracks formed during unloading (Kumamoto et al., 2017) and generate bands of elevated apparent GND density radiating from the indents. Broader zones of elevated GND density, reaching $>10^{14}$ m$^{-2}$, surround each indent. Residual stresses adjacent to the indents exceed 1 GPa and decrease over distances of several micrometers. Residual stress fields also exhibit discontinuities across the traces of microcracks, indicating modification of the stress fields by fracturing.

### *3.3.2 Scanning transmission electron microscopy*

Figure 8 presents STEM images of cross sections through the indents. Figures 8a–c and 8d–f reveal the distributions of dislocations beneath the same indents for which HR-EBSD maps are presented in Figure 7, and which had maximum depths of 300 nm and 800 nm, respectively, in single crystal OP4-2. Beneath both indents, dislocations are contained within a volume approximately 5 μm wide and 4 μm deep. However, the dislocation density beneath the 800-nm indent (Figure 8d and 8e) is higher than that beneath the 300-nm indent (Figure 8a and 8b). Beneath both indents, dislocations are arranged in two sets of linear structures that appear approximately perpendicular to each other in the plane of the section. Based on the measured crystal orientations, one set of dislocations is arranged in a plane that contains the [100] direction, whereas the other set is likely arranged within the (100) plane. The densities of dislocations in these structures are highest near the surface impression of the indent and decrease with distance from the indent, revealing individual dislocations aligned along particular planes (Figures 8a and 8d). Dislocation densities are locally elevated adjacent to intersections of the two sets of dislocations (e.g., bright dislocations near the centre of Figure 8e). Outside the zones of influence of the indents, the crystal is free of dislocations (Figures 8a and 8d).

Figures 8g–i reveal the distribution of dislocations beneath an indent that proceeded to a maximum depth of 600 nm in a crystal containing pre-existing dislocations in sample PI-1488. Dislocations introduced by this indent occupy a volume approximately 3 μm wide and 2 μm deep and lack the orderly structure around the periphery of this zone (Figure 8g) that is evident around the other two indents. Nonetheless, multiple sets of straight dislocations are discernable (Figures 8h and 8i). Where these structures intersect the specimen surface, they produce steps that contribute to the residual impression (Figures 8h and 8i).

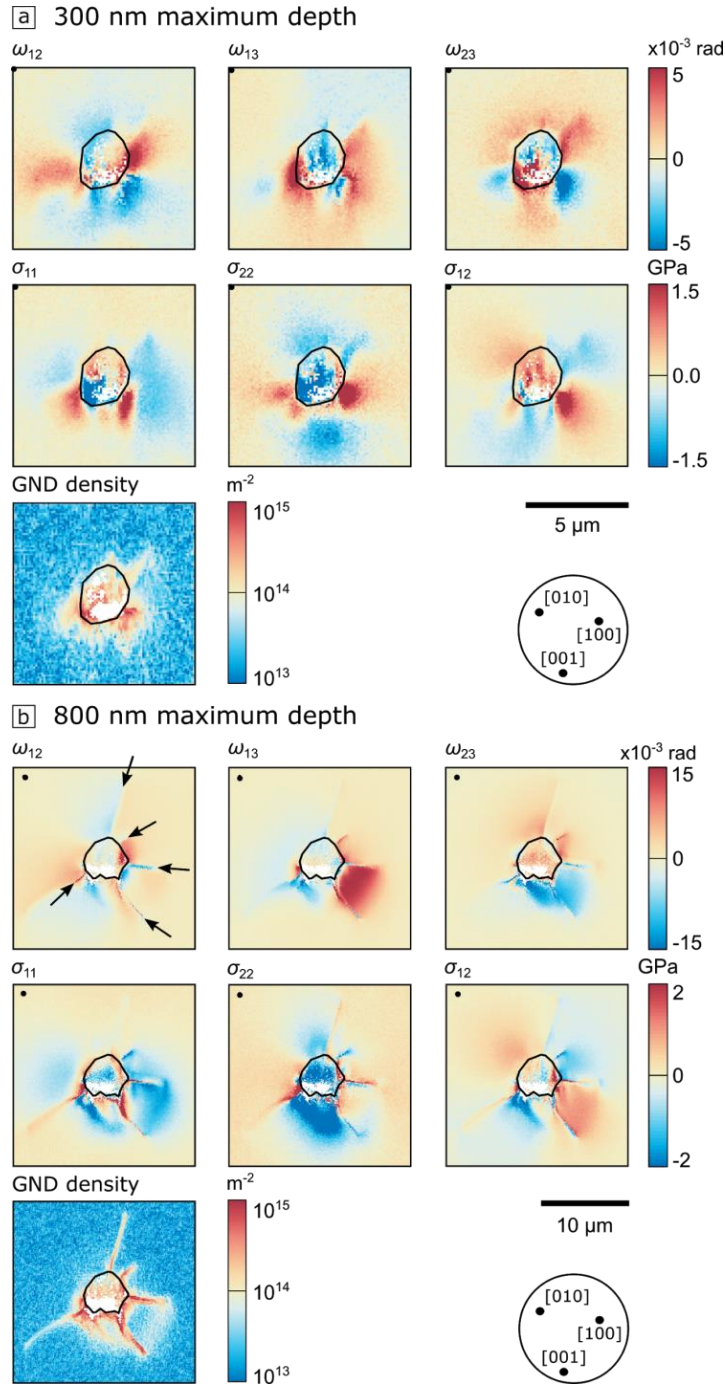

*Figure 7.* HR-EBSD maps of lattice rotations ($\omega_{ij}$), residual stress heterogeneity ($\sigma_{ij}$), and GND density around nanoindents in a single crystal of San Carlos olivine (sample OP4-2 of Kumamoto et al., 2017). a) An indent that reached a maximum depth of 300 nm prior to unloading. b) An indent that reached a maximum depth of 800 nm prior to unloading. Rotations and residual stresses are relative to the crystal orientation and stress state, respectively, at the reference points marked with black dots. Black lines indicate the approximate outlines of the residual impressions. Black arrows in the map of $\omega_{12}$ in (b) mark the tips of the traces of microcracks. Pole figures indicate the crystal orientation and are the same for both indents.

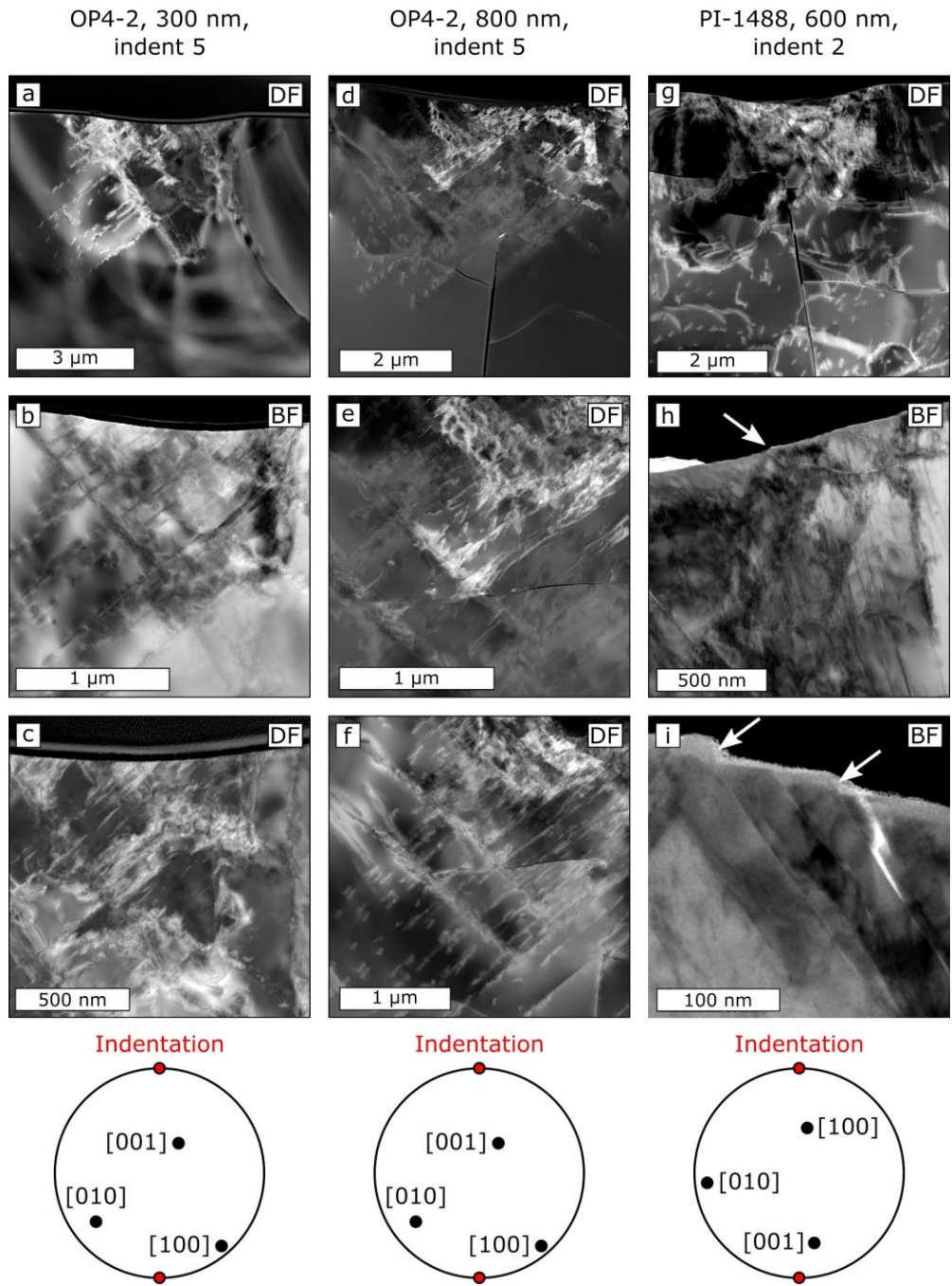

*Figure 8.* STEM images of cross sections through nanoindents. Sample names, maximum indentation depths, and indent numbers are listed at the top of each column. Residual indents are evident as gentle curvature of the specimen surfaces near the tops of the images. *Images in (e) and (f) display the same area at different diffraction conditions, where (e) portrays the change in contrast across the central slip band, while (f) reveals the individual dislocations. Arrows in (h) and (i) indicate steps in the surface of the residual indent. Pole figures indicate the crystal orientation in each column and are in the same reference frame as the STEM images. The crystal orientations of (a–c) and (d–f) are the same as these indents were in the same crystal. HR-EBSD maps of the indents in (a–c) and (d–f) are presented in Figure 7a and 7b, respectively. BF and DF indicate images taken in bright-field and dark-field modes, respectively.*

# 4 Discussion

## *4.1 Deformation mechanisms*

The microstructures demonstrate that deformation occurred by the motion of several types of dislocations. (HR-)EBSD data (Figures 3a, 3d, 3e, and 4a) from the single crystal, San382t, reveal banded substructures with a variety of wavelengths in the range 1–100 μm. The most prominent set of structures have traces approximately parallel to the trace of the (100) plane and orientations within this crystal are dispersed around the [010] axis (Figure 3b). These observations suggest that these structures are slip bands containing dislocations of the (100)[001] slip system. The other set of structures (Figure 3a) have traces approximately parallel to the trace of the (010) plane (Figure 3b) and therefore likely result from slip bands of dislocations on the (010)[100] and/or (010)[001] slip systems. This interpretation is consistent with the orientations of slip bands evident in STEM images of San382t (Figure 6a–c). Similarly, alignment of [010] axes with the compression direction in the aggregate, San382b, also indicates that (010)[100] and/or (010)[001] was the dominant slip system (Figure 3c). Likewise, one set of structures beneath the nanoindents in the undeformed single crystal, OP4-2, are in an orientation indicating that they are slip bands of the (100)[001] slip system, whilst the other set of structures could be slip bands on the (010)[100] and/or (010)[001] systems (Figure 8a–f). Similarly, previous TEM observations and analyses of slip bands expressed as deformation lamellae have documented that the (100)[001] slip system, and particularly [001] screw dislocations, dominate at temperatures < 900°C, with lesser activity of the {110}[001] system (Carter and Ave'lallemant, 1970; Druiventak et al., 2011; Gaboriaud et al., 1981; Idrissi et al., 2016).

## *4.2 Mechanisms of strain hardening and recovery*

The microstructural data reveal several potential strain-hardening mechanisms. Diffraction-contrast fringes in STEM images of dislocations generated in D-DIA experiments suggest that some dislocations are dissociated into partial dislocations bounding stacking faults (Figure 6f and 6h). Poirier (1981) proposed that dislocations on the (100)[001] slip system can split into partial dislocations by the reaction

$$[001] = \frac{1}{12}[013] + \frac{1}{12}[0\bar{1}3] + \frac{1}{12}[013] + \frac{1}{12}[0\bar{1}3], \tag{1}$$

which does not require climb and results in partial dislocations that can all glide on the (100) stacking-fault planes. The TEM observations of Drury (1991) suggest that partial dislocations in olivine can pin other types of gliding dislocations, providing one potential source of hardening.

Details of the microstructures suggest that other short-range interactions among different types of dislocations added resistance to dislocation glide. Some dislocations contain short step-overs or S-shaped bends (sample San377t, Figure 6d–f), similar to those produced by short-range interactions in dislocation dynamics simulations (e.g., Figure 5 of Durinck et al., 2007). Even in the single crystal, San382t, the distribution of dislocations is heterogeneous, with intersecting bands of dislocations resulting in a cellular substructure (Figure 6a–c). The cellular substructure is similar to those produced at temperatures of 600–800°C in the experiments of Phakey et al. (1972) and Druiventak et al. (2012, 2011) and is commonly interpreted to result from local interactions among dislocations during strain hardening in a variety of materials (Kocks and Mecking, 2003). Differences in dislocation density across the intersections of slip bands beneath indents in sample OP4-2 (e.g., Figure 8c and 8f) suggest that interactions between the two types of dislocation impeded their motion. This interpretation is supported by the smaller zone of elevated dislocation density beneath the 600-nm indent in the sample containing pre-existing dislocations (PI-1488, Figure 8a) than beneath the 300-nm indent in the otherwise dislocation-free sample (OP4-2, Figure 8g). Previous detailed TEM observations of single crystals of olivine deformed at temperatures of 600–1090°C indicate that several short-range interactions contribute to strain hardening, including collinear interactions and formation of [101] and [10$\bar{1}$] junctions, along with the formation of dipoles (Mussi et al., 2017; Phakey et al., 1972).

Whilst short-range interactions initially impeded dislocation glide, the observation of the Bauschinger effect by Hansen et al. (2019) indicates that hardening is dominated by long-range elastic interactions. This interpretation is supported by the observation of heterogeneous residual stresses (Figures 4 and 7) that are greater in magnitude than in the starting materials (Figure 5). The residual stresses in the single crystal are particularly informative as they lack a contribution from grain compatibility during decompression and their relationship to the dislocation structures is particularly clear. Bands of residual stress, on the order of 1 μm wide, 1 GPa in magnitude, and alternating in sign, coincide with bands of rotated lattice and elevated GND density (Figure 4a) that mark the locations of slip bands (Figure 6a–c). These stress heterogeneities are characteristic of back stresses between dislocations piled up along their slip planes (e.g., Guo et al., 2014) and indicate a component of kinematic hardening, in which dislocations interact through their long-range stress fields (Kassner et al., 2013). These long-range interactions may contribute to the organisation of the substructure (Montagnat et al., 2006; Wallis et al., 2017), creating wavelengths up to several tens of micrometres in the misorientation associated with slip bands (Figure 3a, 3d, 3e). The stress distributions in the aggregate are broadly consistent with this model but also contain a contribution from stresses imparted by anisotropic expansion of the grains during decompression. Similarly, the nanoindents are surrounded by zones of elevated GND density and stresses up to approximately 1 GPa in magnitude, which have been inferred to control strength during indentation (Kumamoto et al., 2017; Nix and Gao, 1998).

The transition from hardening to constant flow stress in D-DIA experiments indicates the operation of one or more recovery processes (Hansen et al., 2019). Recovery mechanisms involving long-range diffusion, such as dislocation climb, are kinetically inhibited at room temperature. An alternative recovery mechanism is dynamic recovery by annihilation or other rearrangement of dislocations driven by the (local) applied stress (Kocks and Mecking, 2003; Nes, 1997). Cross slip of screw dislocations plays an important role in dynamic recovery of metals (Essmann and Mughrabi, 1979). Phakey et al. (1972) observed evidence of cross slip in olivine deformed at temperatures ≥ 800°C at a pressure of 1 GPa, but not at 600°C. However, it is possible that cross slip was promoted by the higher confining pressures and differential stresses of our experiments (Poirier and Vergobbi, 1978). Similarly, Nes (1997) suggested that dislocation climb may contribute to dynamic recovery, even at low temperatures, if short-range diffusion is driven by locally high stresses. Stress-driven dynamic recovery will generally become more active with the rising applied stress during strain hardening until a constant flow stress is attained.

In summary, a variety of processes contribute to the evolution of strength during low-temperature plasticity of olivine. The similarity in hardening behaviour between single crystals and aggregates (Hansen et al., 2019) indicates that the key processes are predominantly intragranular. Short-range interactions between gliding dislocations and local obstacles, such as dissociated dislocations or full dislocations of another type, provide resistance to dislocation glide that initiates local accumulation of dislocations. Residual-stress heterogeneities with magnitudes on the order of 1 GPa, even in single crystals, indicate that long-range interactions via back stresses between dislocations are the dominant mechanism of strain hardening. Hardening is eventually counteracted by stress-driven dynamic recovery.

## *4.3 Implications for rheological models and lithospheric strength*

The microstructures of samples deformed in D-DIA and nanoindentation experiments provide critical tests of models of low-temperature plasticity. The heterogeneous residual stresses, and their spatial relationship to areas of elevated GND density, in both single crystals and aggregates (Figures 4 and 5), are consistent with the interpretation of Hansen et al. (2019) that strain hardening and the Bauschinger effect result from kinematic hardening induced by long-range dislocation interactions. Similarly, the distributions of dislocations and associated stress fields around nanoindents (Figures 7 and 8) suggest that the same processes control strain hardening observed in the spherical nanoindentation experiments of Kumamoto et al. (2017). The evidence for dislocation interactions influencing dislocation motion beneath the indents is also broadly consistent with models of the indentation size effect, in which yield stress and flow stress are inversely proportional to the size of spherical indenters or the depth of pyramidal indents, respectively (Kalidindi and Pathak, 2008; Nix and Gao, 1998), as demonstrated to occur in olivine by Kumamoto et al. (2017). Those models suggest that smaller deforming volumes result in higher GND densities and stronger interactions among dislocations, which act against the applied stress (Nix and Gao, 1998).

The mechanical data from the D-DIA experiments and the microstructures of the samples deformed at room temperature are similar to those reported previously for deformation at higher temperatures. The portions of stress-strain curves that express strain hardening are broadly similar in shape across temperatures spanning 25–1500°C, but hardening typically occurs over longer strain intervals and is greater in magnitude at lower temperatures (Chopra, 1997; Cooper et al., 2016; Druiventak et al., 2011; Hansen et al., 2019; Hanson and Spetzler, 1994; Post, 1977). Even at high temperatures, strain-hardening 'normal' transients occur in some experiments on single crystals (Cooper et al., 2016; Hanson and Spetzler, 1994), indicating the role of intragranular processes and suggesting potential commonality with the processes documented here to occur at low temperatures. This parallel is supported by the striking similarity between the corrugated lattice formed by slip bands at room temperature (Figures 3 and 4) and microstructures reported by Wallis et al. (2017) for deformation experiments at temperatures of 1000°C and 1200°C, which spanned the transition between exponential and power-law creep. Slip bands in those experiments were also characterised by elevated GND densities that spatially correlate with residual stress heterogeneities of alternating sign. Other similarities between the dislocation types and distributions observed in the present study and those reported to have formed at temperatures of several hundred degrees Celsius have been described above. These similarities in microstructures indicate that, although changes in the mechanism and efficiency of recovery processes may cause changes in mechanical behaviour, the same underlying processes that generate strain hardening may occur across a wide temperature range that extends to, and potentially includes, power-law creep. Testing of this hypothesis against new mechanical data and microstructural observations could provide a microphysical basis for new models of transient creep at high temperatures.

The results of this study support the suggestion of Hansen et al. (2019) that strain hardening is an important consideration for models of lithospheric strength. One example in which such effects may be important is the difference between the strength of the Pacific plate at Hawaii (Zhong and Watts, 2013) and where it enters subduction zones (Hunter and Watts, 2016). The model of Hansen et al. (2019) explains this strength difference in terms of strain hardening due to the accumulation of back stresses between dislocations, with greater strains and therefore hardening where the plate bends into subduction zones. This study demonstrates that the model is consistent with the microstructures of the samples upon which it is based. An additional prediction of the model of Hansen et al. (2019) is that the strength of lithosphere that has undergone strain hardening will be anisotropic. Accumulation of back stress during strain hardening reduces the yield stress during deformation of the opposite sense when the sign of the macroscopic stress is reversed. This effect, the Bauschinger effect, is evident in the mechanical data of Hansen et al. (2019) and is consistent with the stress heterogeneities generated by the dislocation content mapped in this study (Figure 4). Anisotropic lithospheric strength has been inferred from studies of lithospheric flexure in a variety of tectonic settings (Audet and Bürgmann, 2011), although the validity of the methods used to detect it has been questioned (Kirby and Swain, 2014). Commonly, the orientation with the lowest apparent flexural strength is perpendicular to the strike of structures resulting from past compression, i.e., when the bending moment vector is parallel to strike (Audet and Bürgmann, 2011; Audet and Mareschal, 2004; Simons and van der Hilst, 2003). Regardless of the validity of the methods used in flexure studies, the model presented here and by Hansen et al. (2019) predicts that the strength of the lithosphere in deformed regions may be anisotropic in the sense inferred from flexure. A collisional episode will result in accumulation of back stress, which may be retained in the cold, shallow lithosphere. If subsequent flexure induces extension in the shallow lithosphere along the former compression direction, then the apparent yield stress may be reduced relative to flexure in other orientations (Audet and Mareschal, 2004; Simons and van der Hilst, 2003).

Characterisation of the microstructures formed by low-temperature plasticity of olivine, particularly with the novel application of HR-EBSD, provides a new set of diagnostic microstructures with which to identify its past occurrence in natural samples. Specifically, straight dislocations confined within slip bands are associated with stress heterogeneities with magnitudes of several tens of percent of the macroscopic applied stress. Notably, such stress fields provide a microstructural indicator for the occurence of kinematic hardening.

# 5 Conclusions

The microstructures of samples deformed at room temperature in D-DIA and nanoindentation experiments record the microphysical processes of strain hardening during low-temperature plasticity of olivine. Partial dislocations and changes in dislocation density across the intersections of slip bands suggest that short-range interactions impede dislocation glide. However, the presence of intragranular stress heterogeneities with magnitudes on the order of one gigapascal are consistent with aspects of the mechanical data, including the Bauschinger effect (Hansen et al., 2019), indicating that long-range interactions between dislocations via their stress fields are the dominant cause of strain hardening. Similarities between the microstructures formed at room temperature and those formed at higher temperatures suggest that this interpretation may hold for temperatures up to at least several hundred degrees. The results support the flow law of Hansen et al. (2019), which parameterises strain hardening in terms of accumulation of back stress between dislocations and potentially explains differences in the strength of a lithospheric plate in different tectonic settings. The microstructures also provide new criteria to identify the occurrence of low-temperature plasticity and associated strain hardening in natural samples and thereby test the predictions of laboratory-based and geophysical models.

# Acknowledgements

We are grateful for the efficient and excellent technical assistance of Haiyan Chen at beamline 6-BM-B at the Advanced Photon Source. We are also thankful for fabrication of assembly parts by Kurt Leinenweber, Jamie Long, and James King. This research was supported by Natural Environment Research Council grants NE/M000966/1 to LNH, AJW, and DW and 1710DG008/JC4 to LNH and AJW; European Plate Observing System Transnational Access grant EPOS-TNA-MSL 2018-022 to LNH; Advanced Photon Source General User Proposal 55176 to LNH, DLK, DLG, and WBD; and National Science Foundation Awards EAR-1361319 to WD, EAR-1255620 and EAR-1625032 to JMW, and EAR-1806791 to KMK. Data in this paper can be accessed from GFZ Data Services (dataservices.gfz-potsdam.de/portal/).